\documentclass[aps,prl,reprint,superscriptaddress,amsmath,amssymb,floatfix]{revtex4-2}

\usepackage[T1]{fontenc}
\usepackage[utf8]{inputenc}
\usepackage{graphicx}
\usepackage{dcolumn}
\usepackage{bm}
\usepackage{xcolor}
\usepackage{hyperref}
\hypersetup{
    colorlinks=true,
    linkcolor=blue,
    citecolor=blue,
    urlcolor=blue
}

\begin{document}

\title{Spin-momentum locking of polariton edge states in honeycomb lattices}

\author{Andrea Herrero Otermin}
\affiliation{Departamento de F\'isica de Materiales, Universidad Aut\'onoma de Madrid, 28049 Madrid, Spain}

\author{Nicola Carlon Zambon}
\email{nicola.carlonzambon@unipd.it}
\affiliation{Dipartimento di Fisica e Astronomia ``Galileo Galilei'', Universit\`a di Padova, I-35131 Padova, Italy}

\author{Dheerendra Singh}
\affiliation{Department of Physics, Indian Institute of Technology Bombay, Mumbai 400076, India}

\author{Neha Bhoria}
\affiliation{Department of Physics, Indian Institute of Technology Bombay, Mumbai 400076, India}

\author{Rimi Banerjee}
\affiliation{Department of Physics, Indian Institute of Technology Dharwad, Dharwad, Karnataka 580011, India}

\author{Christian G. Mayer}
\affiliation{Julius-Maximilians-Universität Würzburg, Physikalisches Institut and Würzburg-Dresden Cluster of Excellence ctd.qmat, Lehrstuhl für Technische Physik, Am Hubland, 97074 Würzburg, Germany}

\author{Simon Betzold}
\affiliation{Julius-Maximilians-Universität Würzburg, Physikalisches Institut and Würzburg-Dresden Cluster of Excellence ctd.qmat, Lehrstuhl für Technische Physik, Am Hubland, 97074 Würzburg, Germany}

\author{Siddhartha Dam}
\affiliation{Julius-Maximilians-Universität Würzburg, Physikalisches Institut and Würzburg-Dresden Cluster of Excellence ctd.qmat, Lehrstuhl für Technische Physik, Am Hubland, 97074 Würzburg, Germany}

\author{Monika Emmerling}
\affiliation{Julius-Maximilians-Universität Würzburg, Physikalisches Institut and Würzburg-Dresden Cluster of Excellence ctd.qmat, Lehrstuhl für Technische Physik, Am Hubland, 97074 Würzburg, Germany}

\author{Sven H\"ofling}
\affiliation{Julius-Maximilians-Universität Würzburg, Physikalisches Institut and Würzburg-Dresden Cluster of Excellence ctd.qmat, Lehrstuhl für Technische Physik, Am Hubland, 97074 Würzburg, Germany}

\author{Luis Vi\~na}
\affiliation{Departamento de F\'isica de Materiales, Universidad Aut\'onoma de Madrid, 28049 Madrid, Spain}
\affiliation{Instituto Nicol\'as Cabrera, Universidad Aut\'onoma de Madrid, 28049 Madrid, Spain}
\affiliation{Instituto de F\'isica de la Materia Condensada, Universidad Aut\'onoma de Madrid, 28049 Madrid, Spain}

\author{Subhaskar Mandal}
\email{subhaskar@iitb.ac.in}
\affiliation{Department of Physics, Indian Institute of Technology Bombay, Mumbai 400076, India}

\author{Carlos Ant\'on-Solanas}
\email{carlos.anton@uam.es}
\affiliation{Departamento de F\'isica de Materiales, Universidad Aut\'onoma de Madrid, 28049 Madrid, Spain}
\affiliation{Instituto Nicol\'as Cabrera, Universidad Aut\'onoma de Madrid, 28049 Madrid, Spain}
\affiliation{Instituto de F\'isica de la Materia Condensada, Universidad Aut\'onoma de Madrid, 28049 Madrid, Spain}


\begin{abstract}
Transverse-electric/Transverse-magnetic splitting in dielectric-mirror microcavities introduces an effective spin-orbit coupling for photons. While the bulk states remain linearly polarized, for exponentially localized edge states in a photonic lattice, this coupling induces elliptical polarization whose handedness is locked to the propagation direction, analogous to the transverse spin of evanescent electromagnetic waves. We reveal spin-momentum locking through Stokes polarimetry of zigzag edge states in a honeycomb exciton-polariton lattice. The effect persists in a stretched honeycomb supporting a photonic bandgap, where spin-polarized carrier injection enables selective lasing of either chiral edge states. Our results provide a route toward ultrafast spin-controlled unidirectional propagation in polariton systems without external magnetic fields.
\end{abstract}

\maketitle

\begin{figure}[t]
\centering
\includegraphics[width=\linewidth]{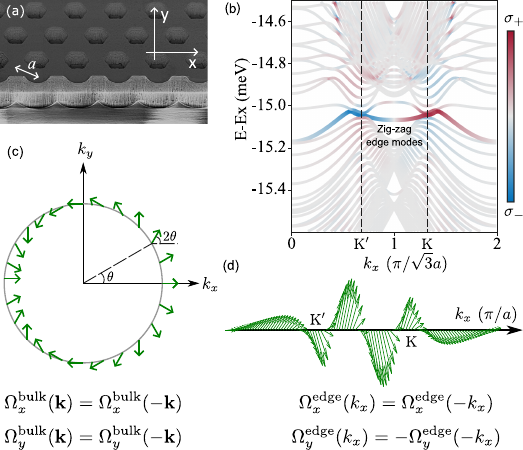}
\caption{\textbf{Spin-momentum locking of zigzag edges states in a honeycomb lattice.} Scanning electron microscopy (SEM) image of a zigzag edge in a honeycomb polariton lattice. The unit-cell size, $a$, is indicated by a double-headed arrow. The red circle marks the size and position of the excitation laser spot. The $x-y$ coordinate axes are shown in the upper-right corner of the panel. (b) Simulation of the projected band structure for a unit cell containing a zigzag edge, the bands encode the circular polarization in a false color scale, showing that the bulk modes remain linearly polarized, while the left- and right-propagating zigzag edge modes exhibit opposite spin polarization. The dashed lines represent K and K$^\prime$ valleys. (c) The green arrows show the distribution of the effective magnetic (proportional to the vector components of $\bm{\Omega}^{\textrm{bulk}}$) field caused by the TE-TM splitting in momentum space in a planar microcavity. (d) Corresponding effective magnetic field (proportional to $\bm{\Omega}^{\textrm{edge}}$) in momentum space ($k_x$) along the ZES.}

\label{fig:sketchspinmomentumlocking}
\end{figure}

\textit{Introduction}---Geometric and symmetry-inspired arguments provide powerful tools for classifying and understanding the physics of condensed-matter systems. These concepts can be extended to structured photonic platforms \cite{Haldane2008}, where the engineering of electric and magnetic material properties has enabled the demonstration of robust unidirectional transport \cite{Wang2009,Zhao2019}, topological lasers \cite{bahari_nonreciprocal_2017,st-jean_lasing_2017,bandres_topological_2018,choi_room_2021,dikopoltsev_topological_2021}, photonic Chern insulators \cite{mittal_photonic_2019,Chenier2026,Roberts2026}, and the observation of Laughlin states of light \cite{schine_synthetic_2016,clark_observation_2020,wang_realization_2024}. Chirality, robustness against disorder, and pseudospin selectivity are highly desirable features for a broad range of photonic and optoelectronic applications.

Unlike electronic systems, photonic platforms lack the Kramers protection afforded by fermionic statistics in spin Hall insulators \cite{Kane2005a,Kane2005b,Bernevig2006}, rendering photonic counter-propagating edge states generally vulnerable to backscattering \cite{lu_topological_2014,Wu2015,ozawa_topological_2019,arora_direct_2021,Rosiek2023,xu_absence_2023,leveque_scattering-matrix_2023}. Consequently, robust unidirectional photonic transport generally requires additional ingredients, such as broken time-reversal symmetry or nonlinear gain effects \cite{Khanikaev2012,bahari_nonreciprocal_2017,Hafezi2013,Yang2020}. 

Since optical gain can be spin-polarized \cite{ando_photon-spin_1998,Hsu2015_UltrafastSpinLaser}, the photon polarization itself provides a natural handle to selectively amplify one of two counter-propagating edge modes \cite{CarlonZambon2019}. Such spin-selective control, however, fundamentally relies on spin-momentum locking (SML), namely a correlation between the polarization state and the propagation direction of light. Remarkably, SML is not exclusively associated with topological photonic systems \cite{widmann_artificial_2026}. Rather, it is a universal property of exponentially localized electromagnetic waves, originating from the complex wavevector of evanescent fields and the constraints imposed by Maxwell's equations \cite{VanMechelen:16,Shi2021,Hosten:2008,Parappurath2020}. 
Exciton-polariton honeycomb lattices provide an ideal platform to investigate this physics \cite{Carusotto2013,kusudo_stochastic_2013,Klembt2018}. These structures realize a photonic analogue of graphene \cite{Jacqmin2014}, while the distributed Bragg mirrors defining the microcavities naturally introduce a polarization-dependent photonic dispersion that acts as an effective spin-orbit coupling for polaritons \cite{Kavokin2005,leyder_observation_2007,Kammann2012,Sala2015, Whittaker2020}. Whereas inversion symmetry enforces linear polarization of the bulk Bloch modes, we show that exponentially localized zigzag edge states (ZES) acquire a finite degree of elliptical polarization whose handedness is intrinsically locked to the propagation direction. In this picture, topology determines the existence of the edge state, whereas its spin texture originates from its evanescent nature. In previous experiments, the SML for ZES was obscured by strain-induced splitting of orthogonal linear polarizations at the interface \cite{Milievi2015}.

Here we experimentally demonstrate that ZES in a polariton honeycomb lattice exhibit intrinsic SML arising solely from their evanescent character. We observe this spin texture through full Stokes polarimetry in a gapless honeycomb lattice. We then engineer a Kekulé-like lattice distortion~\cite{Wu2015}, thereby realizing a topological gapped phase. Exploiting spin-selective optical gain, we finally demonstrate directional edge-state lasing controlled entirely by the pump helicity, providing an all-optical route toward spin-controlled transport in absence of external magnetic fields.

\textit{Samples and setup}---The devices are fabricated from a GaAs-based semiconductor microcavity consisting of a $5\lambda/2$ cavity spacer enclosed by 30- and 35-pair AlAs/Al$_{0.15}$Ga$_{0.85}$As distributed Bragg reflectors. Three groups of four 13-nm-thick GaAs quantum wells are embedded at the antinodes of the intracavity optical field. The sample is held at $10$~K in an open-loop helium cryostat. At cryogenic temperatures, the microcavity operates in the strong exciton-photon coupling regime, giving rise to exciton-polaritons \cite{Weisbuch1992}. Independent measurements of the exciton-photon avoided crossing allow determining the vacuum Rabi splitting $\hbar\Omega_R \approx 8.4~\mathrm{meV}$ \cite{SI}.

The microcavity is laterally patterned into arrays of coupled micropillars forming regular and stretched honeycomb lattices \cite{Wu2015}. The coupling between neighboring pillars is engineered through their spatial overlap, which is controlled by the center-to-center pillar separation. Each unit cell consists of six micropillars arranged in a benzene-like hexagon, with intra- and inter-cell separations given by $V_\mathrm{a}d$ and $V_\mathrm{b}d$, respectively. Here, $d=2.5~\mu$m is the pillar diameter, while the dimensionless parameters $V_\mathrm{a,b}<1$ determine the overlap between adjacent pillars and hence the nearest-neighbor coupling strength~\cite{MichaelisdeVasconcellos2011}. In this work, we investigate a regular honeycomb lattice with $V_\mathrm{a}=V_\mathrm{b}=0.8$ and a stretched honeycomb lattice with $V_\mathrm{a}=0.9$ and $V_\mathrm{b}=0.7$. Figure~\ref{fig:sketchspinmomentumlocking}(a) shows the scaning electron microscopy (SEM) image of a regular honeycomb polariton lattice with a zigzag edge. 

The optical characterization of the samples is performed in a micro-photoluminescence ($\mu$-PL) setup based on a confocal microscope coupled to a spectrometer equipped with a charge-coupled device (CCD) camera. Non-resonant excitation is provided by a continuous-wave Ti:Sapphire laser tuned to the first reflectivity minimum above the distributed Bragg reflector stop band (1.638 eV). The excitation polarization is controlled using a linear polarizer and a quarter-wave plate. The sample photoluminescence is collected through the same microscope objective used for excitation. The near- and far-field emission can be imaged onto the entrance slit of the spectrometer, enabling energy-resolved measurements in real and momentum space and the tomographic reconstruction of the polaritonic band structure \cite{Houdr1994}. Full Stokes polarimetry is implemented in the detection path using a rotating quarter-wave plate followed by a fixed linear polarizer \cite{Wang2006_QWP}. Further details of the experimental setup and polarization analysis are provided in the Supplementary Material \cite{SI}.

\textit{Spin-momentum locking}---The finite photonic stop band of the distributed Bragg reflectors makes the cavity reflectivity depend on both the in-plane wavevector and the polarization of the confined photons. As a result, the microcavity exhibits a momentum-dependent birefringence that is equivalent to an effective in-plane magnetic field acting on the photon pseudo-spin \cite{Kavokin2005},
\begin{equation}\label{eq:Htetm}
    H_{\mathrm{TE\text{-}TM}}\propto
    \Omega_x\sigma_x+\Omega_y\sigma_y,
\end{equation}
where $\sigma_{x,y}$ are the Pauli matrices, $\Omega_x=-\partial_x^2+\partial_y^2$, and $\Omega_y=-2\partial_x\partial_y$. In momentum space, the corresponding effective field exhibits the characteristic double-winding texture shown in Fig.~\ref{fig:sketchspinmomentumlocking}(c) and is invariant under inversion symmetry ($\mathbf{k}\rightarrow-\mathbf{k}$). Consequently, for bulk Bloch modes with real wavevector $(k_x,k_y)$, the TE-TM field only fixes the relative phase between the $\sigma^+$ and $\sigma^-$ polarization components while preserving equal spin populations, resulting in vanishing circular polarization.

This intuition changes for exponentially localized Bloch modes at a lattice edge. Because the edge-state wavefunction decays into the bulk, the momentum component perpendicular to the boundary becomes imaginary, $k_y \rightarrow i\lambda_y$, so that the edge state is described by a complex wavevector. As discussed in Ref.~\cite{VanMechelen:16}, a complex dispersion implies that edge modes necessarily carry a finite transverse spin, whose handedness is locked to the propagation direction. 

To illustrate this mechanism, we compute the projected band structure of a honeycomb ribbon that is periodic along $x$ and finite along $y$, with a zigzag lower edge and a bearded upper edge (see Supplementary Materials \cite{SI}). As shown in Fig.~\ref{fig:sketchspinmomentumlocking}(b), the bulk bands remain linearly polarized, consistent with the symmetry arguments discussed above. In contrast, the ZES appearing between the $K$ and $K'$ valleys acquire a finite degree of circular polarization, while states outside this momentum range are instead localized on the upper bearded edge. The finite dispersion and the breaking of chiral symmetry of the zigzag-edge band originate from next-nearest-neighbor hopping processes \cite{Mangussi2020}.

Within the effective edge-state model described in the Supplementary Materials \cite{SI}, the origin of this polarization can be understood directly from the TE-TM Hamiltonian. Replacing $k_y$ by the evanescent decay constant $i\lambda_y$ yields unequal coupling strengths for the two circular polarization components, proportional to $(k_x+\lambda_y)^2$ and $(k_x-\lambda_y)^2$, respectively. Consequently, the edge states acquire a finite degree of circular polarization,
\begin{equation}\label{eq:SML}
S_3 \propto \frac{2\lambda_y k_x}{k_x^2+\lambda_y^2},
\end{equation}
which changes sign under momentum reversal, $S_3(k_x)=-S_3(-k_x)$. Equivalently, the expectation value of the effective TE-TM field satisfies $\Omega_x^{\mathrm{edge}}(k_x)=\Omega_x^{\mathrm{edge}}(-k_x)$ and $\Omega_y^{\mathrm{edge}}(k_x)=-\Omega_y^{\mathrm{edge}}(-k_x)$, as shown in Fig.~\ref{fig:sketchspinmomentumlocking}(d). Counter-propagating ZES therefore carry opposite circular polarizations, establishing SML.

\begin{figure}[t]
    \centering
    \includegraphics[width=0.98\linewidth]{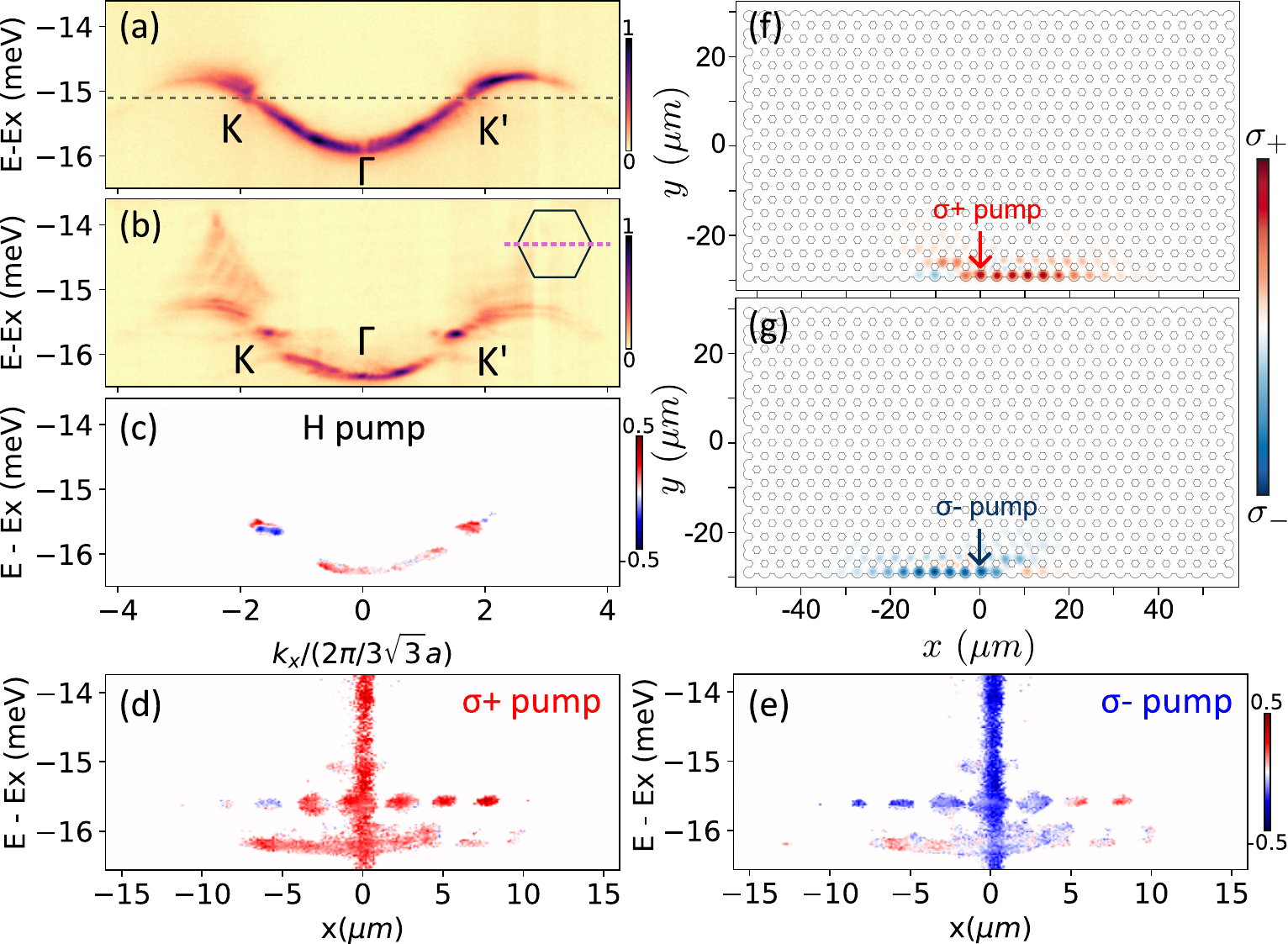}
    \caption{\textbf{Experimental observation of SML in a ZES of a normal honeycomb lattice: linear regime.} (a)/(b) PL intensity map of the dispersion relation of the honeycomb lattice measured in the bulk/along the zigzag edge. The dashed line indicates the energy of the Dirac points. The inset schematically indicates the momentum-space cut across the Brillouin zone. (c) $S_3$ map of the dispersion relation along the zigzag edge under horizontally polarized nonresonant excitation, showing opposite circular polarization for counter-propagating ZES at $K$ and $K'$. (d,e) $S_3$ map of the energy- and real space-resolved emission measured under right- and left-circularly polarized excitation, respectively, revealing SML propagation of the ZES. (f,g) Corresponding Gross--Pitaevskii simulations. All the experimental figures are acquired in the linear regime, under weak non-resonant laser driving, with an excitation spot matching the pillar diameter as indicated in Fig. \ref{fig:sketchspinmomentumlocking}(a).}
    \label{fig:NHC_linear}
\end{figure}

\textit{Results}---We first consider a regular honeycomb lattice ($V_\text{a} = V_\text{b}$). We use a weak, linearly polarized excitation ($P = 0.08~\text{mW}$) addressing a single pillar in the bulk of the lattice to characterize the in-plane polariton dispersion relation. Figure~\ref{fig:NHC_linear}(a) shows a cut of the dispersion relation along the $K$–$\Gamma$–$K'$ direction, as indicated in the inset of Fig.~\ref{fig:NHC_linear}(b). We observe a gapless dispersion that becomes linear in the vicinity of the Dirac cones at $K$ and $K'$. A fit of the dispersion yields nearest- and next-nearest-neighbor coupling strengths of $t = 0.39$ meV and $t' = 0.06$ meV, respectively.
Figures~\ref{fig:NHC_linear}(b,c) show the same dispersion relation, now acquired under identical excitation conditions but with the excitation spot moved to the zigzag edge of the lattice (cf. Fig.~\ref{fig:sketchspinmomentumlocking}). The color code in panels (b) and (c) encodes the intensity ($S_0$) and degree of circular polarization ($S_3$) of the photoluminescence, respectively. At the $K$ and $K'$ points, we observe the emergence of two bright features corresponding to the ZES. The $S_3$ map confirms that ZES with opposite momenta exhibit opposite handedness, with $|S_3| \approx 0.4$.

Next, we evidence the SML effect in real space. Figures~\ref{fig:NHC_linear}(d,e) show the corresponding $S_3$ maps as a function of energy and position along the zigzag edge under $\sigma^+$ and $\sigma^-$ nonresonant excitation, respectively. Under circularly polarized pumping in our platform, carriers can be partially spin-polarized, thereby favoring emission into co-circularly polarized modes \cite{Hsu2015_UltrafastSpinLaser,Lindemann2019}. We observe that a $\sigma^\pm$ pump preferentially enhances the edge state with positive or negative group velocity, respectively, providing a direct visualization of SML. These observations can be also reproduced using a linearized mean-field Gross--Pitaevskii equation, see~\cite{SI} for details. The numerical results, shown in Figs.~\ref{fig:NHC_linear}(f, g), are in good agreement with the experiment and reproduce the SML in ZES under tightly-localized circularly polarized nonresonant excitations. Since triangular honeycomb geometries can be terminated entirely by zigzag edges, the same mechanism enables SML propagation around sharp bends, as shown in the Supplementary Materials \cite{SI}. 

This behavior suggests the possibility of enhancing mode selectivity and hence emission directionality under circular pumping in the nonlinear regime~\cite{Klaas2019,CarlonZambon2019}.
Unfortunately, in the normal honeycomb lattice, the ZES overlap spectrally with bulk modes and therefore cannot be selectively amplified without also feeding the bulk. Consistent with this picture, we could not achieve edge-state lasing in this geometry. We therefore fabricated a stretched honeycomb lattice, whose band gap spectrally isolates the edge modes from the bulk continuum, enabling selective edge-mode amplification.

\begin{figure}[t]
    \centering
    \includegraphics[width=1\linewidth]{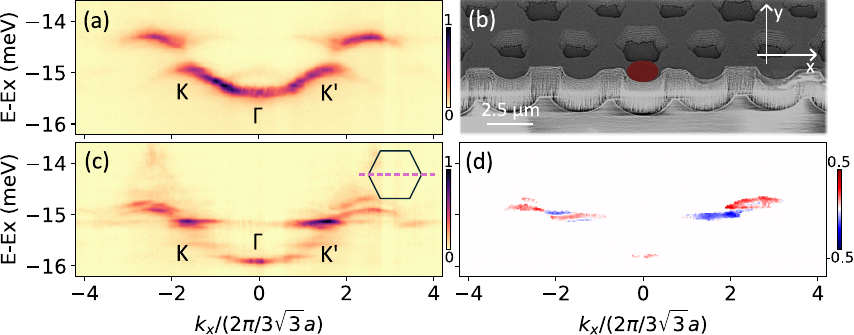}
    \caption{\textbf{Experimental observation of SML in ZES of a stretched honeycomb lattice: linear regime.}  (a) PL intensity map of the dispersion relation of the lattice in the bulk: a gap emerges in the Dirac points. (b) SEM image of the zigzag edge. (c) Dispersion relation measured along the zigzag edge. (d) Corresponding $S_3$ map of the dispersion relation along the zigzag edge under horizontally polarized nonresonant excitation. The SML directionality is reversed with respect to the normal honeycomb lattice. The experimental figures have been acquired in the linear regime ($P{=}0.08$ mW), with a laser spot matching the pillar diameter as indicated in panel (b).}
    \label{fig:SHC_linear}
\end{figure}

Figure~\ref{fig:SHC_linear}(b) shows a SEM image of the stretched honeycomb lattice. In this geometry, the Kekul\'e-O bond ordering folds the original $K$ and $K'$ valleys onto the $\Gamma$ point of the enlarged unit cell. As predicted in Ref.~\cite{Wu2015}, modifying the ratio between intra- and inter-cell couplings opens a band gap while preserving the $C_6$ symmetry. In our structure, the reduced inter-benzene coupling lifts the Dirac degeneracy and gives rise to the gap visible in the bulk dispersion of Fig.~\ref{fig:SHC_linear}(a). Despite the gap opening, the zigzag boundary continues to support edge-localized states, which now lie inside the gap and are spectrally separated from the bulk bands, as shown in Fig.~\ref{fig:SHC_linear}(c).

Next, we resolve the circular polarization of these edge modes. The corresponding $S_3$ map, shown in Fig.~\ref{fig:SHC_linear}(d), reveals that spin-momentum locking persists in the stretched lattice. Interestingly, its sign is reversed relative to the normal honeycomb lattice [Fig.~\ref{fig:NHC_linear}(c)], indicating that the modified coupling landscape alters the underlying pseudospin texture of the zigzag edge states. Although SML arises from the interplay between the TE–TM field and the evanescent ZES, the spin-dependent propagation direction is not determined solely by the topological properties of the dispersion. Since the zigzag-edge band is intrinsically flat, its group velocity is governed by longer-range hopping terms and thus depends on microscopic parameters. Our Gross–Pitaevskii simulations reproduce the experimental observation that the group velocities of the ZES are reversed with respect to those of a conventional honeycomb lattice. A tight-binding analysis further shows that this reversal results from the competition between spin-conserving hoppings and TE--TM-induced spin-flip hopping \cite{SI}.\\

\begin{figure}[t]
    \centering
    \includegraphics[width=1\linewidth]{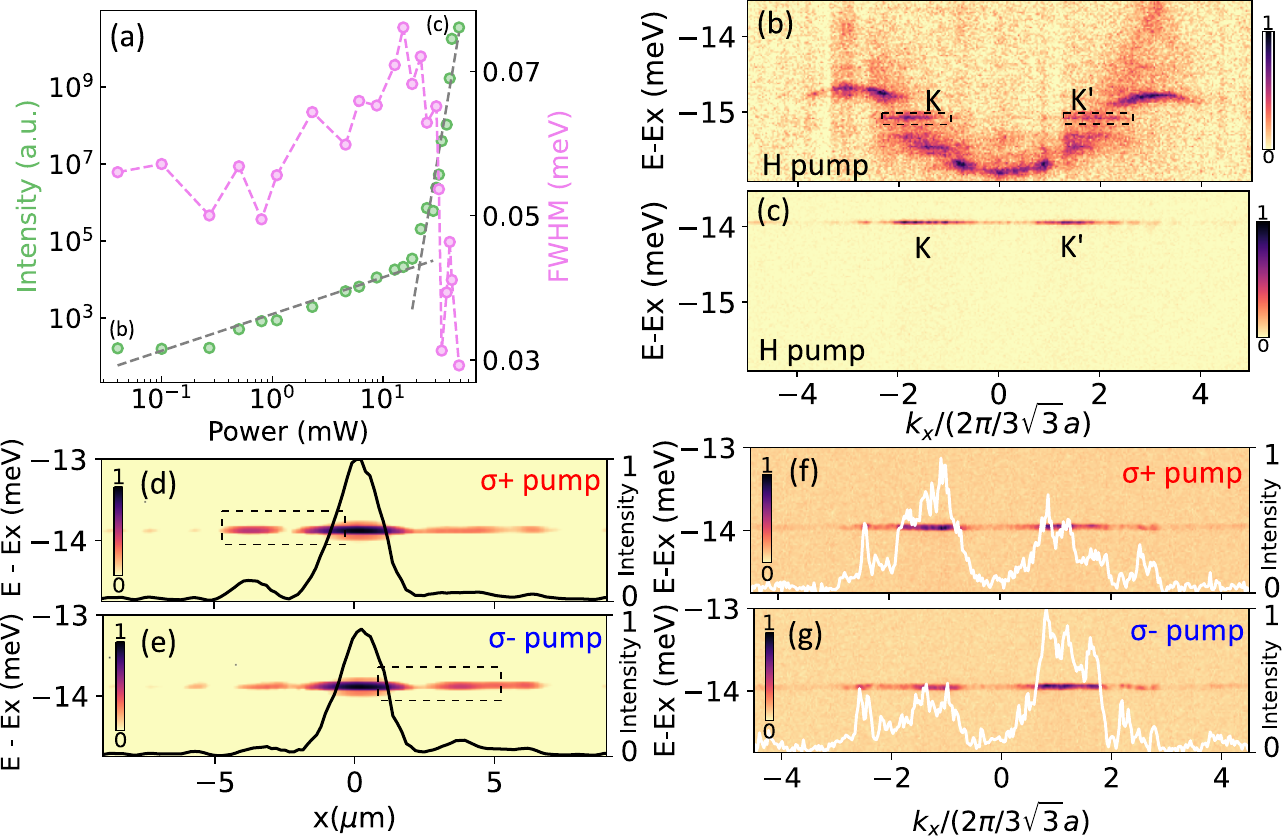}
    \caption{\textbf{SML lasing from ZES in a stretched honeycomb lattice.} (a) Power-dependent emission intensity (green points) showing the nonlinear threshold behavior ($P_{th} = 21.5$ mW, with an elliptical laser spot covering five pillars along the edge) and line narrowing (pink points) associated with polariton lasing in the edge-state modes. (b,c) PL intensity map of the dispersion relation along the ZES for pump powers $P{=}0.005P_{th}/1.95P_{th}$, respectively. (d,e) Spatially and energy-resolved PL intensity map of the ZES under $\sigma^\pm$-circularly polarized excitation, respectively, in the lasing regime ($P{=}1.95P_{th}$). The dashed boxes highlight the propagating direction of the edge-state lasing emission under $\sigma^\pm$ pumping. The black traces in each panel represent the intensity profile versus $x$ at the lasing ZES energy. (f,g) Dispersion relation of the ZES under $\sigma^\pm$-circularly polarized excitation, respectively, in the lasing regime ($P{=}1.74P_{th}$). The white traces in each panel represent the intensity profile versus $k_x$ at the lasing ZES energy.}
    \label{fig:SHC_lasing}
\end{figure}

The spectral isolation provided by the band gap separates the edge states from the bulk modes, enabling us to selectively promote polariton lasing in the edge states. Figure~\ref{fig:SHC_lasing} follows the transition from the linear to the lasing regime. An elongated non-resonant excitation spot covering approximately five edge pillars was used throughout all measurements. Figure~\ref{fig:SHC_lasing}(a) shows the integrated emission intensity and linewidth of the edge mode as a function of pump power. A clear lasing threshold is observed at $P_{\mathrm{th}} = 21.5$~mW, evidenced by the characteristic nonlinear increase in emission intensity and a pronounced linewidth narrowing.

The dispersion relations measured below and above the lasing threshold are shown in Figs.~\ref{fig:SHC_lasing}(b) and (c), respectively, under linearly polarized pumping. Above threshold, the ZES lasing emission dominates the spectrum, exhibiting a linewidth that approaches the spectrometer resolution ($40~\mathrm{\mu eV}$). The SML effect is directly observed in real space in Figs.~\ref{fig:SHC_lasing}(d,e), which show the energy-resolved emission under $\sigma^+$ and $\sigma^-$ nonresonant pumping, respectively. Owing to spin-selective gain, the two excitation helicities preferentially promote lasing in the edge states with opposite $S_3$. Because of spin–momentum locking, changing the helicity of the excitation switches the preferred propagation direction of the lasing edge mode --see dashed boxes in panels (d) and (e)-- demonstrating all-optical control of directional propagation through spin-selective gain. For comparison, the same effect is observed in momentum space in the dispersion relations shown in Figs.~\ref{fig:SHC_lasing}(f,g). The intensity profiles at the ZES lasing energy clearly reveal the SML effect, with predominant emission from the $K$ and $K'$ edge states under $\sigma^+$ and $\sigma^-$ nonresonant pumping, respectively.

Our results show that, in the absence of external perturbations lifting the degeneracy of orthogonal pseudospin modes in polariton lattices, SML emerges naturally for localized edge states at lattice interfaces. Here, we focus on ZES in honeycomb lattices; however, we expect this phenomenon to be a general consequence of the complex wavevector characterizing evanescent edge modes, as predicted both by the minimal model presented in the Supplementary Material and by more general theoretical descriptions \cite{VanMechelen:16}. Remarkably, SML thus arises naturally for the very states that are most relevant for robust routing of optical information, without the need for magnetic fields or synthetic gauge fields \cite{bahari_nonreciprocal_2017,widmann_artificial_2026}. Instead, it originates from the interplay between the photonic analog of spin–orbit coupling \cite{Sala2015,Parappurath2020} and the exponential localization of the edge modes. In the nonlinear regime, we further show that spin-selective gain transfers the polarization of the exciton reservoir to the edge-state population, enabling all-optical control over the directional propagation of spin–momentum-locked polaritons.

Although the relatively short propagation length of the edge modes--limited by their modest group velocity and the finite polariton lifetime--precludes a direct experimental assessment of their robustness against structural disorder or defects, our results suggest that a degree of protection may nevertheless be present. In particular, the spin-polarized reservoir induces different energy shifts for the two counter-propagating ZES, reducing their spectral overlap. This mechanism complements their pseudospin orthogonality, further suppressing back-scattering channels that would otherwise couple the two modes \cite{Yang2020,Real2021_ChiralEmission}.

\textit{Conclusions}---In conclusion, we have experimentally demonstrated SML in the ZES of honeycomb exciton-polariton lattices through polarization-resolved measurements in momentum and real space. We show that the effect originates from the evanescent character of the edge modes and persists in stretched honeycomb lattices supporting a photonic band gap. Furthermore, the combination of spin-selective gain and edge-state lasing enables all-optical control of the propagation direction of chiral polariton modes. Beyond the steady-state regime explored here, time-resolved measurements could provide direct access to the ultrafast formation dynamics of the spin texture and the propagation of spin-polarized wave packets along the lattice edge, opening new opportunities for studying non-equilibrium spin transport in photonic systems.

\begin{acknowledgments}
A.H.O. acknowledges the ``Beca de Colaboración” 2024-2025 from the Ministerio de Educación y Formación Profesional and the ``Becas de Posgrado en el extranjero” from Fundación La Caixa. L.V and C.A-S. acknowledge the support from the projects from the Ministerio de Ciencia e Innovaci\'on PID2023-148061NB-I00 and PCI2024-153425, the “María de Maeztu” Program for Units of Excellence in R\&D (CEX2023-001316-M). C. A.-S. acknowledges the Grant “Leonardo for researchers in Physics 2023” from Fundaci\'on BBVA and the support from the Comunidad de Madrid fund “Atracci\'on de Talento, Mod. 1”, Ref. 2020-T1/IND-19785 and the project from the Ministerio de Ciencia e Innovaci\'on CNS2025-165107. The Würzburg group gratefully acknowledges financial support by the Free State of Bavaria and the Deutsche Forschungsgemeinschaft (DFG, German Research Foundation) through the Würzburg-Dresden Cluster of Excellence ctd.qmat - Complexity, Topology
and Dynamics in Quantum Matter (EXC 2147, project-id 390858490). D. S., N. B., and S. M. acknowledge support from the IIT Bombay IRCC Seed Grant No. RD/0525-IRCCSH0-006. R. B. acknowledges support from the ANRF, Grant No. RJF/2025/000278. We thank Alejandro Gonz\'alez-Tudela for discussions during the early stages of this work.
\end{acknowledgments}

\bibliography{references}

\end{document}